\documentstyle[11pt,newpasp,twoside,epsf]{article}

\def\edcomment#1{\iffalse\marginpar{\raggedright\sl#1\/}\else\relax\fi}
\reversemarginpar

\begin{document}

\title{Long-duration neutron star X-ray transients in quiescence: the
{\itshape Chandra} observation of KS 1731--260} 

\author{Rudy Wijnands} \affil{Center for Space Research, Massachusetts
Institute of Technology, NE80-6055, 77 Massachusetts Avenue,
Cambridge, MA 02139, USA}


\begin{abstract}
I will briefly discuss the implications of the recent quiescent
{\itshape Chandra} observation of the long-duration neutron star
transient KS 1731--260 (Wijnands et al. 2001a) on models for quiescent
systems other than the Brown et al. (1998) cooling neutron star model.
However, the {\it Chandra} results are not very constraining; for
those models to be consistent with the data, one only has to assume
that the system parameters of KS 1731--260 (e.g., the neutron star
spin rate and magnetic field properties; orbital parameters) are very
similar to those of the other systems.  I will also discuss the
available quiescent data of other long-duration neutron star
transients in the context of the Brown et al. (1998) model.
\end{abstract}

\section{Introduction}

X-ray transients spend most of their time in quiescence. The exact
emission mechanisms which produce the quiescent X-rays are not yet
understood. To explain the neutron star emission, several models have
been put forward. So far, the model which assumes that the emission is
due to the cooling of the neutron star (e.g., van Paradijs et
al. 1987; Verbunt et al. 1994; Brown et al. 1998) has been most
successful in reproducing the existing quiescent data and in producing
testable predictions. In the most elaborate version of this model, the
quiescent luminosity of a particular transient depends on the time
averaged accretion rate of this source (Brown et al. 1998).  Recently,
this model could be tested in a new (although anticipated) way. In
early 2001, the neutron star transient KS 1731--260 suddenly became
quiescent again, after having actively accreted for more than a
decade. A {\itshape Chandra} observation was performed on this source
only a few months after this transition to test the Brown et
al. (1998) model (Wijnands et al. 2001a). If the long duration of the
outburst of KS 1731--260 is typical for this source {\it and} its
quiescent time is similar to that of the other neutron star transients
(of order years to a few decades), then the system was detected at a
luminosity (bolometric luminosity of $\sim 2 \times 10^{33}$ erg
s$^{-1}$) and a neutron star temperature ($kT \sim 0.3$ keV) too low
compared to what would have been expected. The implications for the
Brown et al. (1998) model and its validity for KS 1731--260 are
discussed in detail by Wijnands et al. (2001a). If the Brown et
al. (1998) model applies to KS 1731--260, its low luminosity might
indicate that this system is quiescent for several hundreds of
years. This would suggest that several hundreds of such systems might
be present in our Galaxy and they might form a new class of X-ray
transients (see Wijnands et al. 2001a).

\section{Alternative models}

The striking similarities between KS 1731--260 and the other systems,
combined with the extra assumptions and possible adjustments which
have to be made to the Brown et al. (1998) model (Wijnands et
al. 2001a), could be used to argue that the cooling neutron star model
might not be the correct model to explain the quiescent emission of
neutron star transients.  Alternative models include residual
accretion (see, e.g., Campana et al. 1998; Campana \& Stella 2000;
Bildsten \& Rutledge 2000; Narayan et al. 2001 for discussions) or the
models which use a neutron star magnetic field (e.g., Campana et
al. 1998; Campana \& Stella 2000; Robertson \& Leiter 2001\footnote{It
is interesting to note that Robertson \& Leiter (2001) predicted a
luminosity of $\sim10^{33}$ erg s$^{-1}$ for KS 1731--260. Despite
this success, their model was proposed to explain the power-law tail
in the X-ray spectra of quiescent neutron star systems, which is not
the dominate spectral component in KS 1731--260 (Wijnands et
al. 2001a).}). In all these models, it is expected that KS 1731--260
should be very similar to the other systems if its system parameters
(i.e., the neutron star spin rate and the magnetic field properties;
the binary parameters) are very similar to those of the other
systems. The fact that the spin rates of KS 1731--260 and Aql X-1 (as
inferred from the nearly coherent oscillations observed during type-I
X-ray bursts) are very similar (524 Hz versus 549 Hz; Smith et
al. 1997; Zhang et al. 1998), indicates that at least the spin rates
of those systems are very similar. The orbital parameters of KS
1731--260 are not known, but the recent tentative discovery of the
optical counterpart of this system (Wijnands et al. 2001b), might
allow a determination of its orbital parameters. Any small differences
in the details of the quiescent X-ray properties between the systems
(e.g., the exact contribution of the hard power-law tail to the total
0.5--10 keV X-ray luminosity) might simply be due to small differences
in the system parameters.  However, the arguments why such models have
problems explaining the quiescent neutron star emission in general
still apply and it remains to be determined which (if any) of the
models can produce accurately all of the quiescent properties (e.g.,
Campana et al. 1998; Bildsten \& Rutledge 2000; Menou \& McClintock
2001; Narayan et al. 2001).

\section{Other quiescent long-duration neutron star transients}

Besides KS 1731--260, several more systems can be identified as
(possible) long-duration transients. Some of those systems have been
observed in quiescence and although their results are not as clean as
for KS 1731--260, it is useful to compare them with KS 1731--260 and
discuss them in the context of the Brown et al. (1998) model.

\subsection{EXO 0748--676}

This source was discovered with {\it EXOSAT} in February 1985 (Parmar
et al. 1986) but before that (on 22 May 1980) it was serendipitously
observed with {\it EINSTEIN} with a quiescent luminosity of $\sim
10^{34}$ erg s$^{-1}$ (Parmar et al. 1986; Garcia \& Callanan 1999),
at least a factor of ten larger than the other systems. Usually, this
high quiescent luminosity is explained as due to a relatively high
level of residual accretion during this particular observation (see,
e.g., Garcia \& Callanan 1999). However, a different explanation can
be proposed by comparing this system with KS 1731--260. Since its
discovery, EXO 0748--676 has persistently been detected at relatively
high luminosities ($>10^{36}$ erg s$^{-1}$), and, therefore, this
source can be regarded as a long-duration transient which has been
active now for over 15 years (although it is possible we have
witnessed the birth of a new persistent source\footnote{The difference
between a persistent source and a long-duration transient is somewhat
arbitrary because in principle every persistent source turned on in
the past and will turn off in the future.}). If EXO 0748--676 is
typically active {\it and} quiescent for a few decades, then from the
Brown et al. (1998) model, we would expect a luminosity for this
system in quiescence of $10^{34-35}$ erg s$^{-1}$, consistent with
what has been observed.  So, although KS 1731--260 did not behave as
expected (based on the simplest version of the Brown et al. [1998]
model), EXO 0748--676 might behave as expected.  High sensitivity
observations of this latter system are needed when it becomes
quiescent again to study its quiescent spectrum. If the luminosity is
indeed due to the cooling of the neutron star, then its spectrum
should resemble that of the other systems (a black-body-like
spectrum), but with a higher neutron star temperature.  However, the
{\it EINSTEIN} quiescent data of EXO 0748--676 was fitted using a
black-body (although other one-component models fitted equally well)
with a $kT$ of $\sim$0.2 keV (Garcia \& Callahan 1999), which is lower
than expected. But, {\it Chandra} or {\it XMM-Newton} observations are
needed to really constrain its quiescent spectrum.

\subsection{MXB 1659--298}

MXB 1659--298 was discovered in 1976 (Lewin et al. 1976) and had a
clear outburst in 1978 (Lewin et al. 1978). The source was dormant
until April 1999 when it was found to be in outburst again (in 't Zand
et al. 1999). Since then, the source could be detected with the {\it
RXTE} all sky monitor until the writing of this paper (31 July
2001). Therefore, the source has been active for over two years and
might also be considered a long-duration transient (see also Wijnands
et al. 2001a), although two years is relatively short compared to the
11.5 years of KS 1731--260 and the more than 15 years of EXO
0748--676. Verbunt (2001) reported that during a 1991 {\it ROSAT}/PSPC
observation, the source was not detected with an upper limit on the
unabsorbed flux of $1 - 2 \times 10^{-14}$ erg s$^{-1}$ cm$^{-2}$
(0.5--10 keV; estimated with W3PIMMS\footnote{W3PIMMS can be found at
http://heasarc.gsfc.nasa.gov/Tools/w3pimms.html} assuming a black-body
spectrum with $kT\sim0.3$ keV and using the count rate values provided
by Verbunt [2001] but with the updated column density provided by
Oosterbroek et al. [2001]; see also the latter paper for a similar
estimate but for 0.2--2.4 keV). The distance estimates to this source
(as obtained from radius expansion bursts) range from 10 kpc (Muno et
al. 2001) to 13 kpc (Oosterbroek et al. 2001), resulting in a 0.5--10
keV luminosity upper limit of 1--4 $\times 10^{32}$ erg
s$^{-1}$. Assuming that the upper limit on the bolometric luminosity
is only a factor of a few higher (e.g., Rutledge et al. 2000), then
this source had an even lower quiescent luminosity than KS
1731--260. If the time averaged mass accretion rate of the past 25
years in MXB 1659--298 is a good indication of its time averaged mass
accretion rate of the past several thousands of years, then also this
system is too dim in quiescence. It is also possible that the first
two short outbursts are more typical for this source and the long
outburst is a rare phenomenon (note that this is also possible for the
other long-duration transients discussed here, including KS
1731--260). Regardless of the true averaged outburst duration of this
source, high sensitivity observations are needed when it becomes
dormant again to test the effect of the long period of accretion on
the neutron star crust and interior.

\subsection{4U 2129+47}

4U 2192+47 was considered to be a persistent source until in 1983 the
source was not detected with {\it EXOSAT} (Pietsch et al. 1983,
1986). Although the exact beginning of its long active episode is
unknown (so the duration of its active state is unclear), it is
possible that 4U 2192+47 is also a long-duration X-ray transient
(although its possible that we have witnessed the turn off of a
persistent source).  The source was detected in quiescence with the
{\it ROSAT}/HRI at $3\times10^{33}$ erg s$^{-1}$ (0.3--2.4 keV; Garcia
1994) and the {\it ROSAT}/PSPC at $6 \times 10^{32}$ erg s$^{-1}$
(0.5--10 keV; Garcia \& Callanan 1998; the bolometric luminosity would
be about a factor 2 higher; Rutledge et al. 2000). The X-ray spectrum
of the source in the latter data set was consistent with a black-body
with a $kT$ of approximately 0.2 keV. If the outburst and quiescent
duration so far observed for 4U 2192+47 are typical for this source,
then these luminosity and black-body temperature are lower than
expected from the Brown et al. (1998) model. However, during the long
episode during which the source could not be detected at high levels
(since September 1983; Pietsch et al. 1983) and the first detection of
the source in quiescence (December 1991; Garcia 1994), the source
could have cooled down. However, the exact cooling time of neutron
stars is unknown and is model dependent.

\subsection{Globular cluster source X 1732--304 (Terzan 1)}

In the early eighties, {\it Hakucho} detected a bursting X-ray source
in the globular cluster Terzan 1 (Makishima et al. 1981; Inoue et
al. 1981). In 1985, a steady X-ray source was detected (X 1732--304)
consistent with this globular cluster and it is most likely the same
source as the bursting source (Skinner et al. 1987).  Since then, it
has persistently been detected at luminosities between a few times
$10^{35}$ erg s$^{-1}$ and $\sim10^{37}$ erg s$^{-1}$ (see Figure 3 of
Guainazzi et al. 1999). Recently, Guainazzi et al. (1999) reported
that during a {\it BeppoSAX} observation this source could only be
detected at a 2--10 keV luminosity of $1.4\sim10^{33}$ erg s$^{-1}$
(for a distance of 4.5 kpc), with a spectrum consistent with a
black-body spectrum ($kT \sim 0.34$ keV) plus a power law (with photon
index of $\sim$1). These spectral parameters are very similar to those
of the neutron star X-ray transients in quiescence and it is likely
that X 1732--304 became suddenly quiescent after having actively
accreted for over 15 years (see Guainazzi et al. 1998). Thus this
source is a clear example of a long-duration transient, similar to KS
1731--260.

Using the spectral parameters given by Guainazzi et al. (1998), the
quiescent luminosity of X 1732--304 can be converted into a 0.5--10
keV luminosity of $\sim 5 \times 10^{33}$ erg s$^{-1}$. The bolometric
luminosity will be most likely about a factor of two higher
($\sim10^{34}$ erg s$^{-1}$; see, e.g., Rutledge et al. 2000). This is
considerably higher than the bolometric luminosity of KS 1731--260 in
quiescence and might indicate that X 1732--304 behaved more as
expected in quiescence due to its prolonged period of high accretion
(based on the Brown et al. [1998] model). The fact that the {\it
BeppoSAX} quiescent observation of this source was taken at
approximately two years after the last detection of the source above
$\sim10^{35}$ erg s$^{-1}$ (Molkov et al. 2001), demonstrates that the
exact moment when the source became dormant is unknown. The source
could have been in quiescence for about two years before the {\it
BeppoSAX} observation and it could be initially even more luminous and
could have cooled down considerably (this depends on the exact cooling
time of the neutron star crust and interior). However, due to the low
angular resolution of {\it BeppoSAX}, it can also not be excluded that
the detected source might not be X 1732--304 but another
low-luminosity globular cluster source (see also Guainazzi et
al. 1998). In that case, X 1732--304 might be less luminous than
assumed in the above discussion and might be more similar to KS
1731--260. A {\it Chandra} image of Terzan 1 will reveal whether or
not the {\it BeppoSAX} source is indeed X 1732--304 (its position from
a {\it ROSAT}/HRI observation is known to $\sim5''$; Johnston et
al. 1994) and whether the luminosity if contaminated by other nearby
sources. X 1732--304 might be a excellent candidate to test the
cooling neutron star model proposed by Brown et al. (1998).

\section{Conclusions}

The quiescent {\it Chandra} observation of KS 1731--260 can constrain
the cooling neutron star models proposed for the quiescent emission of
neutron star transients. However, it is less constraining for other
types of models. In those models, KS 1731--260 is expected to be very
similar to the other quiescent systems if its orbital parameters
and/or its neutron star parameters (i.e., magnetic field parameters,
spin rate) are very similar to those of the other systems.

It is clear that KS 1731--260 is part of a group of transients which
do not disappear after a few weeks to months but are active for years
to decades. Several such sources can be identified for which
observations were also performed in quiescence. Two of those systems,
EXO 0748--676 and X 1732--304 might indeed be brighter in quiescence
than the other systems, possible due to the heating of the neutron
stars during the long periods of accretion. Three other systems (4U
2129+47, MXB 1659--298, and KS 1731--260) seem to have anomalously low
luminosities in quiescence and might prove to be very constraining for
the models dealing with the heating of the crust and the core of
neutron stars in X-ray binaries. However, only for KS 1731--260 is a
clear picture available about its outburst behavior (i.e., the exact
mass accretion rate during outburst; the outburst duration) {\it and}
was the quiescent observation performed within a few months after the
source become quiescent again. For the other sources, the outburst
behavior is not well known (i.e., its duration; e.g., EXO 0748--676,
MXB 1659--298, 4U 2129+47), the quiescent observation were performed
years after (4U 2129+47, X 1732--304) or before (MXB 1659--298, EXO
0748--676) a long-duration outburst, and source confusion might play a
role (X 1732--304)

\section{Acknowledgement}

I thank Jon Miller for carefully reading this manuscript before
submission. This work was supported by NASA through Chandra
Postdoctoral Fellowship grant number PF9-10010 awarded by CXC, which
is operated by SAO for NASA under contract NAS8-39073.

\end{document}